\documentclass[11pt]{article}
\usepackage{amsthm,enumitem}
\usepackage{amsmath,amsfonts,amssymb,graphicx}
 \usepackage{latexsym}
 \usepackage{mathptmx}
 \usepackage{mathrsfs} 
\usepackage{mathtools}

\usepackage{mathpazo} 
\usepackage[dvipsnames]{xcolor} 
\usepackage{hyperref}
\usepackage{mathrsfs} 
\usepackage{csquotes}

\newtheorem*{theorem*}{Theorem}

\newcommand{\NN}{\mathbb N}

\newcommand{\RR}{\mathbb R}
\newcommand{\CC}{\mathbb C}

\newcommand{\fA}{\mathfrak{A}}

\newcommand{\fK}{\mathfrak{K}}
\newcommand{\fR}{\mathfrak{R}}

\def\cS{{\cal S}}

\def\bK{\boldsymbol K}
\def\bR{\boldsymbol R}
\def\bc{\boldsymbol c}

\newcommand{\be}{\begin{equation}}
\newcommand{\ee}{\end{equation}}

\title{\Large The basic resolvents of position and momentum operators
  form a total set in the resolvent algebra}

\author{Detlev Buchholz${}^{(1)}$ and Teun D.H. van Nuland${}^{(2)}$ \\[2mm]
\small 
${}^{(1)}$ Mathematisches Institut, Universit\"at G\"ottingen, \\
\small Bunsenstr.\ 3-5, 37073 G\"ottingen, Germany \\[5pt]
\small
${}^{(2)}$ Department of Mathematics, University of New South Wales, \\
\small Sydney \ NSW 2052, Australia.
}
\date{}

\begin{document}

\maketitle

\begin{abstract}
  \noindent 
  Let $Q$ and $P$ be the position and momentum operators of a 
  particle in one  
  dimension. It is shown that all compact operators can be approximated
  in norm by linear combinations of the basic resolvents
  $(aQ + bP - i r)^{-1}$
  for real constants $a,b,r \neq 0$. This implies that the 
  basic resolvents form a total set (norm dense span)
  in the C*-algebra $\fR$ generated by the resolvents,
  termed resolvent algebra.
  So the basic resolvents share this property with the
  unitary Weyl operators, which span the Weyl algebra.
   These results obtain for finite systems of particles in any
  number of dimensions. The resolvent algebra of infinite systems
  (quantum fields), being the inductive limit of its finitely
  generated subalgebras, is likewise spanned by its basic resolvents.
\end{abstract}

\section{Introduction}
The resolvent algebra $\fR$ of a single particle on the
real line is the C*-algebra generated by all resolvents
of linear combinations of its position and momentum operators,
\be
 (aQ  + bP - ir )^{-1} \, , \quad a,b \in \RR \, ,
\ r \in \RR \setminus \{ 0 \} \, . 
\ee
It was introduced in \cite{BuGr1} and shown to have 
properties of physical interest that are not shared by the Weyl algebra,
being generated by unitary exponentials of those linear
combinations. Many operators, such as physically significant 
Hamiltonians, are affiliated with the
resolvent algebra; moreover, in contrast to
the Weyl algebra, it is stable under the action of a large
family of dynamics, describing interaction \cite{BuGr1,BuGr2}. 

\medskip 
It is frequently argued
that the Weyl algebra is more convenient for computations of states
since it suffices to determine the states on individual Weyl operators,
making use of the fact that their span generates the algebra.
It is the aim of the present note to show that this feature is also 
shared by the resolvent algebra. Namely the linear span of the basic 
resolvents is norm dense in $\fR$. 

\medskip 
It has been shown in \cite{BuGr1} that the algebra $\fR$ acts faithfully
in the Schr\"o\-dinger representation on $L^2(\RR)$, where
$Q$ acts as a multiplication operator and $P$ by differentiation. We 
restrict our attention in the following to this representation
and make use of the notation 
\be
R_{(a,b)}(r) \coloneqq (a Q + b P - i r)^{-1}  
\, , \quad (a,b) \in \RR^2 \, , \ \ r \in \RR \setminus \{ 0\} \, .
\ee
Thus $R_{(a,b)}(r)^* = R_{(a,b)}(-r)$ and  $R_{(0,0)}(r) = i/r$.

\medskip
Turning to the products of resolvents that are different from
multiples of the identity, $R_{(a,b)}(r) R_{(a',b')}(r') \in \fR$,
there appear two cases. Either $a b' = a' b$. Then the resolvents commute
and are elements of the abelian C*-algebra $\fA_{(a,b)} \subset \fR$
generated by $R_{(a,b)}(r)$ for arbitrary $r \in \RR \setminus \{ 0\}$.
It coincides with the norm closure of the linear
span generated by these
resolvents. This can be seen by noticing that arbitrary products
of the resolvents for different values of $r$ are obtained
by making use of the resolvent equation
\be\label{eq:first resolvent identity}
R_{(a,b)}(r)  R_{(a,b)}(r') = i(r' - r)^{-1}
(R_{(a,b)}(r) -  R_{(a,b)}(r')) \, .
\ee
Thus, since $r \mapsto R_{(a,b)}(r)$ is norm continuous   
for $r \neq 0$, the linear combinations of these resolvents are norm
dense in $\fA_{(a,b)}$, as stated. 

\medskip
The second possibility is $ab' \neq a'b$. Then 
$R_{(a,b)}(r) R_{(a',b')}(r') \in \fR$ is a compact operator on $L^2(\RR)$ 
since the underlying generators are canonically conjugate,
cf.~\cite[Thm.\ XI.20]{ReSi}. 
The C*-algebra $\fK$ of compact operators does not contain any
finite linear combinations of basic resolvents, however, apart from  $0$.
A total set of compacts is obtained by weak integrals of the
functions $(a,b) \mapsto R_ {(a,b)}(r)$ with test functions
$f \in \cS(\RR^2)$. 
Since the resolvents are discontinuous in norm,
it is not clear from the outset that these integrals can be approximated
in norm by linear combinations
of resolvents. That this is possible will be shown in
the subsequent section, thereby establishing our main result.
\begin{theorem*}
  The resolvent algebra $\fR$ coincides with the norm closure of the
  span of its generators,
  \be
  \fR = \overline{\text{\rm span}} \, 
  \{ R_{(a,b)}(r) : (a,b) \in \RR^2 \, ,
  \ r \in \RR \setminus \{ 0 \} \} \, .
  \ee
\end{theorem*}

We restrict our attention here to a single particle
in one spatial dimension. But, as outlined in the concluding
remarks, our results extend to finite and infinite systems in any
number of dimensions.
They complement and simplify an argument, alluded to in \cite{vN23},
that is based on analogous results for the classical theory in \cite{vN23}
and on Berezin quantization \cite{vN19}.

\section{Proof of the theorem}
The missing step in the proof of the theorem is the demonstration that
compact operators can be approximated in norm by linear combinations of
the basic resolvents. Since $a,b \mapsto R_{(a,b)}(r)$ is continuous
in the strong operator topology, we can proceed to the integrals
\be
\overline{R}_f(r) \coloneqq \int \! da db \, f(a,b) R_{(a,b)}(r) \, ,
\quad f \in \cS(\RR^2) \, , \ \ r \in \RR \backslash \{ 0 \} \, .
\ee
By arguments in \cite[Lem.~3.2]{Bu}, they can be shown to
define compact operators. We need to show here that these integrals
can be approximated in norm by linear combinations of the basic resolvents
$R_{(a,b)}(r)$ and that one obtains in this manner a dense set
of compact operators.

\medskip 
Since $\| \overline{R}_f(r) \| \leq \| f \|_1 / |r|$, where
$\| f \|_1$ denotes the norm of $f$ in $L^1(\RR^2)$, 
it suffices to consider the dense set of test functions $f$
with compact support.
We can then approximate the integrals by Riemann sums of the form
\be
\overline{R}_{f,n}(r) \coloneqq
\frac{1}{n^2} \sum_{l,m = -nL}^{nL} f(\tfrac{l}{n}, \tfrac{m}{n}) \, 
R_{(\frac{l}{n}, \frac{m}{n})}(r) \, , \quad n \in \NN \, ,
\ee
where $L \in \NN$ is sufficiently large such that the support of
$f$ is contained in a square about $0$ of side length $2L$. These
sums converge for large $n$ in the strong operator topology
to $\overline{R}_f(r)$.  Now
\begin{align}
& \overline{R}_{f, n}(r)^* \overline{R}_{f, n}(r)  \nonumber \\
& = \frac{1}{n^4}
  \sum_{l,m,l',m' = -nL}^{nL} \delta_{\, lm', l'm} \, f(\tfrac{l}{n}, \tfrac{m}{n})^* 
  f(\tfrac{l'}{n}, \tfrac{m'}{n}) R_{(\frac{l}{n}, \frac{m}{n})}(r)^*
  R_{(\frac{l'}{n}, \frac{m'}{n})}(r) + C_n \, ,
\end{align}
where $C_n \in \fK$ is a sum involving
products of non-commuting resolvents, hence of compact operators,
$n \in \NN$. An estimate of the first term yields
\be
\| \overline{R}_{f, n}(r)^* \overline{R}_{f, n}(r) - C_n \|
\leq \frac{(2nL + 1)^3}{n^4 r^2} \, \| f \|_\infty^2 \, , \quad n \in \NN \, , 
\ee
where $\| f \|_\infty$ denotes the supremum norm of $f$. 
It follows from this estimate by standard arguments
(making use of continuity properties of the square root and the polar
decomposition of operators)
that there is a sequence of compacts $K_n \in \fK $ such
that $\|\overline{R}_{f, n}(r) - K_n \| \rightarrow 0$
in the limit of large $n \in \NN$. Thus these compacts also 
converge in this limit to the operator
$\overline{R}_f(r) \in \fK$ in the strong operator topology.

\medskip 
The dual space of the Banach space $\fK$ consists of normal functionals
$K \mapsto \mbox{Tr} \, K \tau$, where $\tau$ are trace class operators. 
Since the bounded sequence $K_n \in \fK$, $n \in \NN$, converges
in the strong operator topology to $\overline{R}_f(r) \in \fK$,
it is also convergent in the weak Banach space topology of $\fK$. 
This puts us into the position to apply Mazur's lemma \cite[p.~6]{EkTe}.
Namely, there exist convex combinations of these compact operators,
\be
\bK_{n} \coloneqq \sum_{l = n}^{N_n} c_{l,n} \, K_{l} \, ,
\quad c_{l,n} \geq 0 \, , \ \ \sum_{l = n}^{N_n} c_{l,n} = 1 \, , \ \
n \in \NN  \, , 
\ee
which approximate $\overline{R}_f(r)$ in norm, 
$\lim_n \| \bK_{n} - \overline{R}_f(r) \| = 0$. 
We proceed now to the corresponding sums of the basic resolvents,
\be
\overline{\bR}_{f, n}(r) \coloneqq
\sum_{l = n}^{N_n} c_{l,n} \overline{R}_{f, l}(r) \, ,
\quad c_{l,n} \geq 0 \, , \ \ \sum_{l = n}^{N_n} c_{l,n} = 1 \, , \ \
n \in \NN \, .
\ee
It follows that
\begin{align}
\| \overline{\bR}_{f, n}(r) - \overline{R}_f(r) \| & \leq
  \| \overline{\bR}_{f, n}(r) - \bK_{n} \| +
   \| \bK_{n}  - \overline{R}_f(r) \| \nonumber \\
   & \leq \sup_{l \geq n} \| \overline{R}_{f, \, l}(r) - K_l \| +
   \| \bK_{n}  - \overline{R}_f(r) \| \, . 
\end{align} 
So the linear combinations of the basic resolvents entering in   
$\overline{\bR}_{f, n}(r)$ converge in norm
to the compact operator $\overline{R}_f(r)$ in the limit of large
$n \in \NN$.

\medskip
The proof that one obtains in this manner a dense set of
compact operators is accomplished by showing 
that suitable linear combinations of the compact operators $\overline{R}_f(r)$
for $r \in \RR \setminus \{ 0 \}$
approximate in norm the smoothed Weyl operators
\be
\overline{W}_f \coloneqq \int \! da db \, f(a,b) \, e^{-i(aQ + bP)} \, ,
\quad f \in \cS(\RR^2) \, .
\ee
From a well-known theorem of von~Neumann \cite{vNe} it follows that 
the span of the latter operators forms a dense
subalgebra of the compacts $\fK$. 
These operators can be approximated in the strong operator
topology by the sequence 
\be
\overline{W}_{f,n} \coloneqq
\int \! da db \, f(a,b) \, (1 + (i/n)(a Q + bP))^{-n} \, ,
\quad n \in \NN \, .
\ee
The operators $(1 + (i/n)(a Q + bP))^{-n}$, $n \in \NN$, are elements of the
abelian algebra $\fA_{(a.b)}$. As was shown in the preceding section,
cf.\ relation \eqref{eq:first resolvent identity},
they can be approximated in norm
by linear combinations of the basic resolvents $R_{(a,b)}(r)$ for
$r \in \RR \backslash \{ 0 \}$. This 
approximation is uniform in $(a,b) \in \RR^2$.
It implies that the operators
$\overline{W}_{f,n}$ can be approximated in norm by 
linear combinations of the compact operators $\overline{R}_{f}(r)$,
$r \in \RR \backslash \{ 0 \}$; whence, they are also elements of
$\fK$. Since the sequence
$\overline{W}_{f,n} \in \fK$, $n \in \NN$, is convergent in the
strong operator topology to $\overline{W }_f \in \fK$, 
it follows from Mazur's lemma that suitable convex combinations of these
operators converge in norm. Hence $\overline{W }_f$
is the norm limit of linear combinations of the basic
resolvents, completing the proof of the theorem. 

\vspace*{-1mm}
\section{Concluding remarks}
Restricting attention to the simple example of a single particle in one
dimension, we have shown that the corresponding
resolvent algebra coincides with 
the norm closure of the span of the underlying basic resolvents.
As in case of the Weyl algebra, this simplifies the computation
of states on this algebra.

\medskip
Let us briefly illustrate this point
for quasifree states on $\fR$ \cite[Sect.\ 4]{BuGr1}.
In the case at hand, these are determined
by suitable scalar products on $\CC^2$. For vectors
$\bc = (c_1,c_2) \in \CC^2$ they are of the form 
\be
\langle \bc, \bc \rangle = \alpha |c_1|^2 + \beta |c_2|^2 +
(i/2)(c_1^* c_2 - c_1 c_2^*) \, ,
\ee
where $\alpha, \beta > 0$ and $\alpha \beta \geq 1/4$.
According to the preceding results, the corresponding
quasifree states $\omega$ on $\fR$ are fixed by
linear extension of the expectation values of the basic resolvents.
They are given by 
\be
\omega(R_{(a,b)}(r)) = i \, \text{sign}(r) \int_0^\infty \! dt \,
e^{-t |r|} e^{-(t^2/2)  \langle \bc, \, \bc \rangle} \, , 
\ee
where $\bc = (a,b) \in \RR^2$ and $r \in \RR \backslash \{ 0 \}$. 

\medskip
The preceding arguments can be extended to arbitrary finite and infinite
systems in any number of dimensions. As is outlined below,   
the linear span of the basic resolvents is 
norm dense in the resolvent algebras of any finite quantum system.
Since for infinite systems
(quantum fields) the corresponding resolvent algebras
are the C*-inductive limit of nets of such finite subalgebras \cite{BuGr1},
the span of the underlying basic resolvents is norm dense
in those cases as well.

\medskip 
We briefly indicate the straightforward proof for canonically
conjugate position and momentum operators $(Q_m, P_m)$ 
that commute for different  values
of \mbox{$m = 1, \dots , n$}. The corresponding resolvent algebra
$\fR_n$ on $L^2(\RR^n)$ is the C*-algebra generated by the resolvents 
\be
\big(\sum_{m = 1}^n (a_m Q_m + b_m P_m) -ir \big)^{-1} \, , \quad
a_m, b_m \in \RR \, , \ m = 1, \dots , n \, , \ \ r \in \RR \setminus
\{ 0 \} \, .
\ee
The essential step is the demonstration that the product of any
two such resolvents can be approximated in norm by linear combinations
of basic resolvents.
The proof that the linear span of all resolvents is norm dense
in $\fR_n$ is then obtained by an obvious inductive procedure. There are
the following three possibilities. 

\medskip \noindent
(1) \ The generators $G \coloneqq \sum_{m = 1}^n (a_m Q_m + b_m P_m)$
of the two resolvents differ by a constant factor. Then the product
of the resolvents is defined by the resolvent equation, as discussed 
in the introduction.

\medskip \noindent
(2) \ The generators $G_1, G_2$ of the two resolvents commute, but
are not proportional. In that case one can show that the product
of the respective resolvents can be approximated in norm by linear
combinations of the resolvents of $(a G_1 + b G_2)$ with
$(a,b) \in \RR^2$. Proceeding to the joint spectral resolution
of the generators, it amounts to proving that the functions
\be
\RR^2 \ni (x,y) \mapsto (x -i r)^{-1} (y -i r')^{-1} \, ,
\quad r, r' \in \RR \setminus \{ 0 \} \, , \ee
being elements of $C_0(\RR^2)$, can be approximated
in the supremum norm by the span of the basic resolvents 
\be
(x,y) \mapsto (a x + b y - i s)^{-1} \, ,
\quad (a, b) \in \RR^2 \, , \ \ s \in \RR \backslash \{ 0 \} \, .
\ee
Similarly to the case of compact operators, discussed in the
preceding section, this is accomplished by integrating the
basic resolvents with regard to the parameters
$(a, b)$
with test functions $f \in \cS(\RR^2)$. The resulting integrals
are elements of $C_0(\RR^2)$. They can be 
approximated by Riemann sums which converge to the integrals
in the weak topology
of $C_0(\RR^2)$, recalling that its dual
space consists of finite complex measures. Note that the re\-solvents
are continuous with regard to $(a, b)$, uniformly
for $(x,y)$ varying in compact subsets of $\RR^2$. By
applying again the procedure used in case of compact operators
(passage to squares of 
Riemann sums, determination of the contributions in
$C_0(\RR^2)$, estimates of the remainders,
and Mazur's lemma), it follows that convex
combinations of the Riemann sums approximate the integrals
in norm.

\medskip 
That one obtains by this procedure 
a norm dense set of elements of $C_0(\RR^2)$ can be seen
by noting that the Fourier transforms of $f \in \cS(\RR^2)$,
\be
x,y \mapsto \hat{f}(x,y) = \int \! da db \, f(a,b) \, e^{-i(ax +by)} \, ,
\ee
can be approximated in the weak topology of $C_0(\RR^2)$
by linear combinations of integrals involving
the basic resolvents. As before, one
replaces the exponential function by the sequence
$a,b \mapsto (1 + (i/n)(ax +by))^{-n}$, $n \in \NN$.
The latter functions are norm 
limits of finite linear combinations of the basic
resolvents $x,y \mapsto (ax +by -is)^{-1}$, $s \in \RR \backslash \{ 0 \}$, 
uniformly for $(a,b) \in \RR^2$.
It then follows by another application of 
Mazur's lemma that linear combinations
of the basic resolvents approximate the Fourier transforms
of $f \in \cS(\RR^2)$ in the supremum norm. So their span is dense in
$C_0(\RR^2)$ and the product of the resolvents
of $G_1, G_2$ can be approximated in this manner.

\medskip \noindent
(3) If the generators $G_1, G_2$ do not commute, their
commutator is a multiple of the imaginary unit $i$. 
The C*-algebra generated by the basic resolvents of all
linear combinations of $G_1, G_2$ is then isomorphic to the
C*-algebra~$\fR$ of a one-dimensional system, as discussed in the
preceding sections. The product of the resolvents of
$G_1$ and $G_2$ is contained in its compact ideal, so
it can also be approximated in norm by linear
combinations of the basic resolvents.

\medskip
Thus all finite and infinite
resolvent algebras are generated by the span of the
underlying basic resolvents. 
We hope that these results will help to make progress in the
structural analysis of the resolvent algebras, such as
the determination of their automorphism groups and of   
the affiliated operators. 

\vspace*{7mm} \noindent
\textbf{\Large Acknowledgments}

\medskip \noindent 
We acknowledge stimulating discussions with Hendrik Grundling.
DB is grateful to Dorothea Bahns and the Mathematics Institute
of the University of G\"ottingen for their continuing hospitality. 
TvN was supported by ARC grant \ FL17010005.

\end{document}